\documentclass[letterpaper]{article}
\usepackage{amsmath}
\usepackage{amssymb}
\usepackage{mathrsfs}
\usepackage{multirow}
\usepackage{graphicx}
\usepackage[colorlinks,linkcolor=blue,anchorcolor=blue,citecolor=blue,urlcolor=blue]{hyperref}

\begin{document}
\title{\textbf{Analytical description of high-aperture STED resolution with 0-2$\pi$ vortex phase modulation}}
\date{}

\author{Hao Xie$^{1,4}$, Yujia Liu$^{1,2,3}$, Dayong Jin$^3$, and Peng Xi$^{1*}$}
\maketitle
\begin{center}
{ $^1$Department of Biomedical Engineering, College of Engineering, Peking University, Beijing 100871, China;\\
 $^2$ School of Life Sciences and Biotechnology, Shanghai Jiao Tong University, No. 800 Dongchuan Road, Shanghai 200240, China;\\
$^3$Advanced Cytometry Labs, MQphotonics Research Centre, Macquarie University, NSW 2109, Sydney, Australia;\\
$^4$Wallace H Coulter Department of Biomedical Engineering, Georgia Institute of Technology and Emory University, Atlanta, USA\\
 \url{*Email: xipeng@pku.edu.cn}
}
\end{center}
\section*{Abstract}
 Stimulated emission depletion (STED) can achieve optical super-resolution, with the optical diffraction limit broken by the suppression on the periphery of the fluorescent focal spot.
Previously, it is generally experimentally accepted that there exists an inverse square root relationship with the STED power and the resolution, yet without strict analytical description. In this paper, we have analytically verified the relationship between the STED power and the achievable resolution from vector optical theory for the widely used 0-2$\pi$ vortex phase modulation. Electromagnetic fields of the focal region of a high numerical aperture objective are calculated and approximated into polynomials, and analytical expression of resolution as a function of the STED intensity has been derived. As a result, the resolution can be estimated directly from the measurement of the saturation power of the dye and the STED power applied.
\newpage
\section*{Introduction}
In 1873, Ernst Abbe derived the diffraction limit of conventional microscopy as ${{d}_{Abbe}}={\lambda }/{2NA}$, which is also regarded as the resolution limit for optical microscopy ~\cite{abbe1873beitrage}. It was not until the past few decades with the development of super-resolution microscopy techniques that this limit has been surpassed ~\cite{ding2011hacking}. One of these techniques, stimulated emission depletion (STED), belongs to the optical nanoscopy family of reversible saturable optical fluorescence transitions (RESOLFT) ~\cite{keller2007efficient}, in which the point spread function (PSF) of the fluorescence is engineered to obtain a higher resolution which breaks the optical diffraction limit ~\cite{rittweger2009sted}. In particular, a spatially modulated depletion beam is spatially overlapped on the excitation focal spot to effectively remove the peripheral fluorescence distribution with either STED ~\cite{hell1994breaking}, or by exciting the electron to a triplet state (ground state depletion, GSD) ~\cite{folling2008fluorescence}, or through excitation state absorption (ESA) to excite the electron to a higher energy level~\cite{irvine2008direct}. To achieve super-resolution in the lateral plane, a widely used method is to place a 0-2$\pi$ vortex phase modulator in the optical path of the circular polarized depletion beam ~\cite{galiani2012strategies,liu2012achieving,tzeng2011superresolution,Vicidomini2013Nanoscopy}.

To estimate the theoretical super-resolution power of STED, the most widely used expression is the square root law in which the full-width at half-maximum $FWHM=h_0/\sqrt{1+I/I_s}$ which was first developed by Westphal and Hell ~\cite{westphal2005nanoscale} and later by Harke et al. ~\cite{harke2008resolution}. Although partially explains the experimental results, it is an empirical theory without solid theoretical foundation, as the depletion intensity distributions discussed in previous work are not derived from the diffraction theory. The scalar solution to the depletion intensity of a 0-2$\pi$ phase plate has been derived by Watanbe et al. ~\cite{watanabe2004formation}. Yet without considering the vectorial nature of light, such theory can only be applied with a small aperture objective where the paraxial approximation is applicable. As a result, this theory fails for objectives with large apertures. On the other hand, it is possible to numerically simulate the intensity distribution at the focal spot, without analytical expression ~\cite{deng2010effects,hao2010effects}. So far, despite the huge progress in the experimental instrumentation and application of STED in optical nanoscopy, accurate estimation of the performance of a STED system under a known depletion power is still lacking. It is therefore of great importance to establish a new formula describing the maximum achievable resolution of STED optical nanoscopy, based on the relationship between the power of the depletion beam and the measured power when the detected fluorescence intensity is depleted to its half.

In this paper we use a vectorial theory to investigate the laser field in the neighborhood of focus. We present the diffraction analysis in Results, where analytical excitation and depletion light intensity distributions are given and then approximated with polynomials. It is demonstrated that only the chirality of the circular-polarization matching the direction of the 0-2$\pi$ vortex phase plate can generate a central zero doughnut intensity distribution. Furthermore, with a second-order approximation, expressions of FWHM with different saturation functions are derived from system parameters in both CW-STED and pulsed-STED. The previous resolution law $FWHM=h_0/\sqrt{1+I/I_s}$ is validated, and the relationship between the depletion power and the half fluorescence inhibition power are developed. Then we numerically simulate the electromagnetic fields, and compare its results with our analytical results. Finally we discuss the error induced by our approximation, and the scope and limitation of the application to this theory.

\section*{Results}
Wolf et al. derived the vectorial theory of diffraction under asymptotic and the optical rays approximation ~\cite{wolf1959electromagnetic,richards1959electromagnetic}. For a linear polarized excitation laser beam, the electric beam in objective space can be written as:
\begin{equation}
\left.
\begin{aligned}
e_x&=-\frac{i A}{\pi}\int_0^\alpha\int^{2\pi}_0 \cos^\frac{1}{2}\theta sin\theta\{\cos\theta+(1+\cos\theta)\sin^2\phi\}e^{i kr_p\cos\epsilon}d\theta  d\phi \\
e_y&=\frac{ i A}{\pi}\int_0^\alpha\int^{2\pi}_0 \cos^\frac{1}{2}\theta \sin\theta(1-\cos\theta)\sin\phi \cos\phi  e^{ i kr_p\cos\epsilon} d\theta  d\phi\\
e_z&=\frac{ i A}{\pi}\int_0^\alpha\int^{2\pi}_0 \cos^\frac{1}{2}\theta \sin^2\theta \cos\phi  e^{ i kr_p\cos\epsilon} d\theta d\phi
\end{aligned}
\right\}\label{eq:fomalism}
\end{equation}
The coordinates are illustrated in Fig. \ref{fig:schematic}. In an idealistic Gaussian image system,  the above expression yields:
\begin{equation}
\left.
\begin{aligned}
e_x(P)&=- i A_{1x}(I_0+I_2cos2\phi_p)\\
e_y(P)&=- i A_{1x}I_2sin2\phi_p\\
e_z(P)&=-2A_{1x}I_1cos\phi_p
\end{aligned}
\right.\nonumber
\end{equation}
where $A_{1x}=k_1fe_{1x}/2$, $k_1=2\pi/\lambda_{ex}$ and $e_{1x}$ is the linear excitation electric field along the x-axis, and on focal plane integration $I_0$, $I_1$, $I_2$ are given by
\begin{equation}
\left.
\begin{aligned}
I_0&=\int_0^\alpha \cos^\frac{1}{2}\theta \sin\theta(1+\cos\theta)J_0(k_1\rho_p \sin\theta)\exp(ik_1z_p\cos\theta) d\theta \\
  I_1&=\int_0^\alpha \cos^\frac{1}{2}\theta \sin^2\theta J_1(k_1\rho_p \sin\theta)\exp(ik_1z_p\cos\theta) d\theta \\
  I_2&=\int_0^\alpha \cos^\frac{1}{2}\theta \sin\theta (1-\cos\theta) J_2(k_1\rho_p \sin\theta) \exp(ik_1z_pcos\theta) d\theta
\end{aligned}
\right. \nonumber
\end{equation}
where $\rho_p=(x_p^2+y_p^2)^{1/2}$. In STED a 0-2$\pi$ phase mask is generally used, as it can generate the most efficient inhibition pattern for fluorescence depletion ~\cite{keller2007efficient}. The electric field of a depletion beam with distinct polarization states can be calculated  ~\cite{deng2010effects}. Notice $\int^{2\pi}_0 \exp\{ i n\phi+\rho \cos\phi\} d\phi=2\pi i^n J_n(\rho)$, and substitute $\exp( i kr_p\cos\epsilon)= e^{ i kz_p \cos\theta}e^{ i k\rho_p \sin\theta \cos(\phi-\phi_p)} e^{ i k\Phi(\theta,\phi)}$ into Eq. (\ref{eq:fomalism}), where the phase function of 0-2$\pi$ phase plate is $ e^{ i k\Phi(\theta,\phi)}= e^{ i\phi}$, we derive:
\begin{equation}
\left.
\begin{aligned}
e_x&=A_{2x}[\mathscr{I}_1\exp( i\phi_p)-\frac{1}{2}\mathscr{I}_2\exp(- i\phi_p)+\frac{1}{2}\mathscr{I}_3\exp( i3\phi_p)]\\
e_y&=-\frac{ i A_{2x}}{2}[\mathscr{I}_2\exp(- i\phi_p)+\mathscr{I}_3\exp( i 3\phi_p)]\\
e_z&= i A_{2x}[\mathscr{I}_4-\mathscr{I}_5\exp( i 2\phi_p)]
\end{aligned}
\right\}\label{eq:depletion2}
\end{equation}
Here $A_{2x}=k_2fe_{2x}/2$ is proportional to linear depletion electric field, and $k_2=2\pi/\lambda_{de}$. The angular integrations $\mathscr{I}_1-\mathscr{I}_5$ are defined as
\begin{equation}
\left.
\begin{aligned}
\mathscr{I}_1&=\int_0^\alpha  d\theta \cos^\frac{1}{2}\theta \sin\theta J_1(k_2\rho_p \sin\theta)(1+\cos\theta)\exp(ik_1z_p\cos\theta)\\
\mathscr{I}_2&=\int_0^\alpha  d\theta \cos^\frac{1}{2}\theta \sin\theta J_1(k_2\rho_p \sin\theta)(1-\cos\theta)\exp(ik_1z_p\cos\theta)\\
\mathscr{I}_3&=\int_0^\alpha  d\theta \cos^\frac{1}{2}\theta \sin\theta J_3(k_2\rho_p \sin\theta)(1-\cos\theta)\exp(ik_1z_p\cos\theta)\\
\mathscr{I}_4&=\int_0^\alpha  d\theta \cos^\frac{1}{2}\theta \sin\theta J_0(k_2\rho_p \sin\theta)\sin\theta\exp(ik_1z_p\cos\theta)\\
\mathscr{I}_5&=\int_0^\alpha  d\theta \cos^\frac{1}{2}\theta \sin\theta J_2(k_2\rho_p \sin\theta)\sin\theta\exp(ik_1z_p\cos\theta)
\end{aligned}
\right.\nonumber
\end{equation}

Comparison of the 3D intensity profile with result of PSF Lab ~\cite{nasse2010realistic} can be seen in Fig. S1.

For circular polarization incidence, the electric field can be obtained by coordinate rotation ~\cite{voort19893}. The electromagnetic field now consists of contributions of two linear polarized incident fields along the x and y axis:
\begin{equation}
\left.
\begin{aligned}
  E_x(r_p,\theta_p,\phi_p)=e_x(r_p,\theta_p,\phi_p)-e'_{y'}(r'_p,\theta'_p,\phi'_p) \\
  E_y(r_p,\theta_p,\phi_p)=e_y(r_p,\theta_p,\phi_p)+e'_{x'}(r'_p,\theta'_p,\phi'_p)\\
  E_z(r_p,\theta_p,\phi_p)=e_z(r_p,\theta_p,\phi_p)+e'_{z'}(r'_p,\theta'_p,\phi'_p)
\end{aligned}
\right\}\label{eq:totalfield}
\end{equation}
where $r'_p=r_p$,$\theta'_p=\theta_p$,$\phi'_p=\phi_p-\frac{\pi}{2}$.Then the intensity of laser field could be calculated by $I=E_xE_x^*+E_yE_y^*+E_zE_z^*$ for both the excitation and the depletion beam, and residual intensity is $I_{residual}=I_{ex}*\eta(I_{de})$, where $\eta(I_{de})=1/(1+I_{de}/I_s)$ for continuous beams and $\eta(I_{de})=\exp[-(\ln2) I_{de}/I_s]$ for pulsed STED ~\cite{leutenegger2010analytical}.

Now consider the light intensity distribution on focal plane $z_p=0$. Since $J_\nu(x)=\Sigma_{k=0}^{\infty}\frac{(-1)^ k}{k!\Gamma(k+\nu+1)}(\frac{x}{2})^{2k+\nu}$, the approximation to the second-order can be used: $J_0(x)=1-\frac{x^2}{4}$, $J_1(x)=\frac{x}{2}$, and $J_2(x)=\frac{x^2}{8}$.
Define $s=\sin^2\alpha$ and $B(s,p,q)=\int_0^s t^{p-1}(1-t)^{q-1}d t=2\int_0^\alpha \sin^{2p-1}\theta\cos^{2q-1}\theta d\theta$, which is the incomplete Beta function, we can express $I_1-\mathscr{I}_5$ as polynomials of $\rho_p$ with their coefficients given by $a_{01}-b_5$, and their coefficients are incomplete Beta functions of $s$, as shown in Supplementary Information. For a linear or elliptically polarized laser, it is always possible to rotate coordinates to make the polarization a standard ellipse with $A_{1x}=C_1$, $A_{1y}= i C_2$, $A_{2x}=D_1$, $A_{2y}=i D_2$, where $C_1$,$C_2$,$D_1$,$D_2$ are real. Then the excitation and STED doughnut intensities can be expressed as:
\begin{equation}
\left.
\begin{aligned}
I_{ex}(\rho_p,\phi_p)&=(C_1^2+C_2^2)a_{01}^2+2[(C_1^2+C_2^2)(a_{01}a_{02}+a_1^2)\\
&+(C_1^2-C_2^2)(a_{01}a_2+a_1^2)\cos2\phi_p](k_1\rho_p)^2\\
I_{dep}(\rho_p,\phi_p)&=(D_1-D_2)^2b_{41}^2+(D_1^2+D_2^2)b_1^2(k_2\rho_p)^2\\
&+(D_1-D_2)^2(\frac{1}{2}b_2^2+ 2b_{41}b_{42})(k_2\rho_p)^2 \\
&-(b_1b_2+2b_{41}b_5)(D_1^2-D_2^2)\cos2\phi_p(k_2\rho_p)^2
\end{aligned}
\right\}\label{eq:excitationanddepletion}
\end{equation}

This expression gives an analytical description of the excitation and depletion intensities in the vicinity of the focus. A special case is that the depletion  beam is left-hand circularly polarized (the same as the direction of the vortex direction) with $C_1=C_2=\frac{\sqrt{2}}{2}C$ and $D_1=D_2=\frac{\sqrt{2}}{2}D$, which indicates that the center of donut has zero intensity. Then $I_{ex}=C^2[a_{01}^2+2(a_{01}a_{02}+a_1^2)(k_1\rho)^2]$ and $I_{dep}=D_1^2b_1^2(k_2\rho)^2$. 
For an elliptically polarized laser or linearly polarized laser ~\cite{chu2012single}, $D_1\neq D_2$ leads to $I_{de}\neq 0$ at the focus. This is derived from second-order approximation, but still holds true for a strict analytical solution because all the higher-orders annihilate at $r=0$. The results obtained with our methods can also be validated with ~\cite{hao2010effects}, as shown in Fig. S2. The full-width at half-maximum (FWHM) of STED PSF could be estimated accordingly, with the decay functions for CW or pulse cases described in SI.

\section*{Discussion}
The intensity $I_s$ could be measured as the following procedures:
(1) measure the peak fluorescence intensity of excitation beam (here the size of the fluorophore is assumed to be very small, i. e. a delta function);
(2) introduce the depletion beam without phase modulation, whose PSF should be overlapped with the PSF of the excitation;
(3) adjust the depletion power until the peak fluorescence intensity is depleted to its half ~\cite{harke2008resolution}. As a result, the peak depletion intensity is defined as $I_s$. When the CW saturation functions $\eta(I_{de})=1/(1+I_{de}/I_s)$  is considered, to its second-order ~\cite{leutenegger2010analytical}:
\begin{equation}
\left.
\begin{aligned}
FWHM&=\frac{\lambda_{ex}}{\alpha NA\sqrt{1+\beta I_{dep}/I_s}}
\end{aligned}
\right.\label{eq:FWHM}
\end{equation}
where $I_{dep}$ is the non-modulated peak intensity of the depletion beam focus, and $\alpha=\pi\sqrt{-4(a_{01}a_{02}+a_1^2)}/a_{01}s$, $\beta=\frac{b_1^2}{2(a_{01}a_{02}+a_1^2)}
(\frac{\lambda_{ex}}{\lambda_{de}})^2$. For $NA=1.4$ and $n=1.5$, $\lambda_{ex}=635 nm$, $\lambda_{de}=760 nm$, this expression yields $FWHM=\lambda_{ex}/(2.15NA\sqrt{1+0.168I_{dep}/I_s})$. For pulsed-STED with saturation function $\eta(I)=\exp(-ln 2 I/I_s)$, it is more convenient to rewrite excitation into an exponential function ~\cite{leutenegger2010analytical} and thus $\alpha=\pi\sqrt{-2(a_{01}a_{02}+a_1^2)/(ln 2)}/a_{01}s$, $\beta=\frac{(ln 2)b_1^2}{2(a_{01}a_{02}+a_1^2)}(\frac{\lambda_{ex}}{\lambda_{de}})^2$. With the same system parameters, this expression yields $FWHM=\lambda_{ex}(1+0.233I_{dep}/I_s)^{-1/2}/1.83NA$. We found the inverse root law still holds in vectorial theory except for modification factors $\alpha$ and $\beta$, which are incomplete beta functions with aperture angle as their parameters. The relationship between the depletion intensity ratio $I/I_s$ and the fluorescence intensity ratio $P_det/P_0$ is shown in Fig. S4.

In experiment, the measurement of intensity is often replaced with measuring of the laser power, and then calculating the intensity distribution for modulated and unmodulated STED PSFs ~\cite{willig2007sted}. Stimulated emission intensity $I_s$ is often defined as the intensity of fluorescence dropped into its half ~\cite{harke2008resolution,leutenegger2010analytical}, or 1/e ~\cite{westphal2005nanoscale,donnert2006macromolecular}. The resolution of STED is often described with the relationship of $P/\mathscr{P}_s$, where $P$ is the depletion power and $\mathscr{P}_s$ is defined by the depletion power when the detected fluorescence intensity is reduced by half, if we overlap the non-phase modulated depletion PSF with the excitation PSF ~\cite{willig2007sted,harke2008resolution}.

 We used STED3D to simulate the electromagnetic field distributions around the focus. In Fig. \ref{fig:Fig4} we plotted the excitation (a) and the depletion (b) intensities on x-axis of the focal plane. In (a) the simulation intensity was plotted in red, the polynomial approximation of Eq. (\ref{eq:excitationanddepletion}) was plotted in blue, and the exponential approximation was plotted in green. In Fig. \ref{fig:Fig4}(b) the numerical solution to the rigid EM field intensity was plotted in red, and polynomial approximation of Eq. (\ref{eq:excitationanddepletion}) was plotted in blue. As expected, approximations agreed well with the numerical solution of the EM field in the vicinity of focus, but diverged with the increase of the radius.
Then we plotted in Fig. \ref{fig:Fig5} the simulated FWHM (red) and the calculated results in Eq. (\ref{eq:FWHM}) (blue). Figure \ref{fig:Fig5}(a) is the CW case with fraction saturation function and Figure \ref{fig:Fig5}(b) is the pulsed case with exponential saturation function ~\cite{leutenegger2010analytical}. In both figures the two curves overlap better with the increase of depletion intensity.

In above discussion we followed Wolf's diffraction theory and Harke's approximation to derive an analytical resolution expression. Yet, it is difficult to measure the focal intensity of a high-NA objective. Practically, the intensity is calculated with the measurement of the laser power and diffraction distribution ~\cite{harke2008resolution,willig2007sted,vicidomini2011sharper}. First the total fluorescence power with excitation beam is record by a confocal point detector ~\cite{harke2008resolution,willig2007sted}, and then depletion beam is introduced without phase modulation. Reduced detected fluorescence power is recorded with respect of the depletion power, and the exponential-fitted saturation power is defined as $\mathscr{P}_s$ ~\cite{vicidomini2011sharper}. Therefore, it is of great importance to develop relationship between the STED resolution and the STED power~\cite{vicidomini2011sharper}. Numerical evaluation of the factor reveals $P=\mathscr{P}_s$ approximately corresponds to $I=2I_s$ (Appendix Fig. 6). Then the resolution vs. STED power relationship becomes ~\cite{willig2007sted,hell1992properties}
\begin{equation}
FWHM=\frac{\lambda_{ex}}{\alpha NA\sqrt{1+\chi P/\mathscr{P}_s}}
\end{equation}
 where $P$ is the depletion power, $\mathscr{P}_s$ is the saturation power when the fluorescence is reduced by half. Our representation of resolution then becomes $FWHM=\lambda_{ex}/(2.15NA\sqrt{1+0.4P/\mathscr{P}_s})$ and $FWHM=\lambda_{ex}/(1.82NA\sqrt{1+0.5P/\mathscr{P}_s})$ for CW-STED and pulsed-STED, respectively. As an example typical values of parameter $\chi$ for different dyes are listed in Supplementary Information Tab. S1. 

 In this paper we have applied a polynomial approximation of the Bessel integrals, to yield an analytical expression of STED with vectorial waves. The results were validated with previous reported numerical simulations of the electromagnetic wave. The relationship of $1/\sqrt{1+I/I_s}$ has been validated in both CW and pulsed STED cases. We have given new sets of resolution estimation equations, to predict the resolution  achievable for a STED system: the expression of resolution, incident depletion intensity and half-peak-fluorescence depletion intensity is analytically derived with incomplete Beta functions, and the relationship of half-peak-fluorescence depletion intensity and half-total-fluorescence depletion power was established by simulation. The result can be extended to other RESOLFT type super-resolution techniques where the fluorescence point spread function is modulated with another doughnut shape point spread function through 0-2$\pi$ phase modulation.

In the derivation of Eq. (\ref{eq:FWHM}), asymptotic expansion at $r=0$ makes the approximation only available in the vicinity of the focus. In Fig. \ref{fig:Fig4} we found that the curves of analytical and simulation laser intensity overlap well when $r$ was small but diverged with the increase of $r$. In Fig. \ref{fig:Fig4} we found that when $r<80 nm$, the approximated excitation and depletion intensity will induce errors of $5\%$ and $8\%$, respectively. As a result, our theory applies to high aperture objective lenses with $NA>1$, or those with high depletion intensity, roughly when the FWHM is less than $\lambda/3$. For smaller aperture objective lens, Watanabe's theory~\cite{watanabe2004formation} with scalar diffraction theory ~\cite{born1999principles,goodman1968fourier} can provide an acceptable estimation. For the same reason, in Fig. \ref{fig:Fig5} the analytical FWHM curves better predicts results with error less than $5\%$ when $I>10 I_s$. With the development of the high efficiency fluorescent dye and through application of larger STED power, STED resolution of 6 $nm$ has been demonstrated ~\cite{rittweger2009sted}, and typical STED system resolution is $<80 nm$. As usually the best STED nanoscopy resolution at a certain depletion power is concerned, rather than the resolution at moderate power, our result has  a wide scope of application in the accurate estimation of the STED (or RESOLFT) system performance.

There are several improvements that can be made in the future. One is introducing the Gaussian spatial distribution of the incident lasers, which can be done by inserting a Gaussian factor in the integration of $I_0 - \mathscr{I}_5$. Since the beam is generally expanded for overfilling the back aperture of the objective, the planar wave can serve as a very good approximation to the experimental situation. Another possible improvement would result from retaining higher orders in the Bessel summation. It may reduce the error when the radius increases, at the cost of much complexity in expression.


\begin{thebibliography}{10}
\expandafter\ifx\csname url\endcsname\relax
  \def\url#1{\texttt{#1}}\fi
\expandafter\ifx\csname urlprefix\endcsname\relax\def\urlprefix{URL }\fi
\providecommand{\bibinfo}[2]{#2}
\providecommand{\eprint}[2][]{\url{#2}}

\bibitem{abbe1873beitrage}
\bibinfo{}{Abbe, E.}
\newblock \bibinfo{}{Beitr{\"a}ge zur theorie des mikroskops und der
  mikroskopischen wahrnehmung}.
\newblock \emph{\bibinfo{}{Archiv f{\"u}r mikroskopische Anatomie}}
  \textbf{\bibinfo{}{9}}, \bibinfo{}{413--418} (\bibinfo{}{1873}).

\bibitem{ding2011hacking}
\bibinfo{}{Ding, Y.}, \bibinfo{}{Xi, P.} \& \bibinfo{}{Ren, Q.}
\newblock \bibinfo{}{Hacking the optical diffraction limit: Review on recent
  developments of fluorescence nanoscopy}.
\newblock \emph{\bibinfo{}{Chinese Science Bulletin}} \textbf{\bibinfo{}{56}},
  \bibinfo{}{1857--1876} (\bibinfo{}{2011}).

\bibitem{keller2007efficient}
\bibinfo{}{Keller, J.}, \bibinfo{}{Sch{\"o}nle, A.} \& \bibinfo{}{Hell, S.}
\newblock \bibinfo{}{Efficient fluorescence inhibition patterns for resolft
  microscopy}.
\newblock \emph{\bibinfo{}{Opt. Express}} \textbf{\bibinfo{}{15}},
  \bibinfo{}{3361--3371} (\bibinfo{}{2007}).

\bibitem{rittweger2009sted}
\bibinfo{}{Rittweger, E.}, \bibinfo{}{Han, K.}, \bibinfo{}{Irvine, S.},
  \bibinfo{}{Eggeling, C.} \& \bibinfo{}{Hell, S.}
\newblock \bibinfo{}{Sted microscopy reveals crystal colour centres with
  nanometric resolution}.
\newblock \emph{\bibinfo{}{Nature Photonics}} \textbf{\bibinfo{}{3}},
  \bibinfo{}{144--147} (\bibinfo{}{2009}).

\bibitem{hell1994breaking}
\bibinfo{}{Hell, S.} \& \bibinfo{}{Wichmann, J.}
\newblock \bibinfo{}{Breaking the diffraction resolution limit by stimulated
  emission: stimulated-emission-depletion fluorescence microscopy}.
\newblock \emph{\bibinfo{}{Optics Letters}} \textbf{\bibinfo{}{19}},
  \bibinfo{}{780--782} (\bibinfo{}{1994}).

\bibitem{folling2008fluorescence}
\bibinfo{}{F{\"o}lling, J.} \emph{et~al.}
\newblock \bibinfo{}{Fluorescence nanoscopy by ground-state depletion and
  single-molecule return}.
\newblock \emph{\bibinfo{}{Nature Methods}} \textbf{\bibinfo{}{5}},
  \bibinfo{}{943--945} (\bibinfo{}{2008}).

\bibitem{irvine2008direct}
\bibinfo{}{Irvine, S.}, \bibinfo{}{Staudt, T.}, \bibinfo{}{Rittweger, E.},
  \bibinfo{}{Engelhardt, J.} \& \bibinfo{}{Hell, S.}
\newblock \bibinfo{}{Direct light-driven modulation of luminescence from
  mn-doped znse quantum dots}.
\newblock \emph{\bibinfo{}{Angewandte Chemie}} \textbf{\bibinfo{}{120}},
  \bibinfo{}{2725--2728} (\bibinfo{}{2008}).

\bibitem{galiani2012strategies}
\bibinfo{}{Galiani, S.} \emph{et~al.}
\newblock \bibinfo{}{Strategies to maximize the performance of a sted
  microscope}.
\newblock \emph{\bibinfo{}{Optics Express}} \textbf{\bibinfo{}{20}},
  \bibinfo{}{7362--7374} (\bibinfo{}{2012}).

\bibitem{liu2012achieving}
\bibinfo{}{Liu, Y.} \emph{et~al.}
\newblock \bibinfo{}{Achieving $\lambda$/10 resolution cw sted nanoscopy with a
  ti: Sapphire oscillator}.
\newblock \emph{\bibinfo{}{PLoS ONE}} \textbf{\bibinfo{}{7}},
  \bibinfo{}{e40003} (\bibinfo{}{2012}).

\bibitem{tzeng2011superresolution}
\bibinfo{}{Tzeng, Y.} \emph{et~al.}
\newblock \bibinfo{}{Superresolution imaging of albumin-conjugated fluorescent
  nanodiamonds in cells by stimulated emission depletion}.
\newblock \emph{\bibinfo{}{Angewandte Chemie International Edition}}
  \textbf{\bibinfo{}{50}}, \bibinfo{}{2262--2265} (\bibinfo{}{2011}).

\bibitem{Vicidomini2013Nanoscopy}
\bibinfo{}{Vicidomini, G.} \emph{et~al.}
\newblock \bibinfo{}{Sted nanoscopy with time-gated detection: Theoretical and
  experimental aspects}.
\newblock \emph{\bibinfo{}{PLoS ONE}} \textbf{\bibinfo{}{8}},
  \bibinfo{}{e54421} (\bibinfo{}{2013}).

\bibitem{westphal2005nanoscale}
\bibinfo{}{Westphal, V.} \& \bibinfo{}{Hell, S.}
\newblock \bibinfo{}{Nanoscale resolution in the focal plane of an optical
  microscope}.
\newblock \emph{\bibinfo{}{Physical Review Letters}} \textbf{\bibinfo{}{94}},
  \bibinfo{}{143903} (\bibinfo{}{2005}).

\bibitem{harke2008resolution}
\bibinfo{}{Harke, B.} \emph{et~al.}
\newblock \bibinfo{}{Resolution scaling in sted microscopy}.
\newblock \emph{\bibinfo{}{Opt. Express}} \textbf{\bibinfo{}{16}},
  \bibinfo{}{4154--4162} (\bibinfo{}{2008}).

\bibitem{watanabe2004formation}
\bibinfo{}{Watanabe, T.} \emph{et~al.}
\newblock \bibinfo{}{Formation of a doughnut laser beam for super-resolving
  microscopy using a phase spatial light modulator}.
\newblock \emph{\bibinfo{}{Optical Engineering}} \textbf{\bibinfo{}{43}},
  \bibinfo{}{1136} (\bibinfo{}{2004}).

\bibitem{deng2010effects}
\bibinfo{}{Deng, S.}, \bibinfo{}{Liu, L.}, \bibinfo{}{Cheng, Y.},
  \bibinfo{}{Li, R.} \& \bibinfo{}{Xu, Z.}
\newblock \bibinfo{}{Effects of primary aberrations on the fluorescence
  depletion patterns of sted microscopy}.
\newblock \emph{\bibinfo{}{Optics Express}} \textbf{\bibinfo{}{18}},
  \bibinfo{}{1657--1666} (\bibinfo{}{2010}).

\bibitem{hao2010effects}
\bibinfo{}{Hao, X.}, \bibinfo{}{Kuang, C.}, \bibinfo{}{Wang, T.} \&
  \bibinfo{}{Liu, X.}
\newblock \bibinfo{}{Effects of polarization on the de-excitation dark focal
  spot in sted microscopy}.
\newblock \emph{\bibinfo{}{Journal of Optics}} \textbf{\bibinfo{}{12}},
  \bibinfo{}{115707} (\bibinfo{}{2010}).

\bibitem{wolf1959electromagnetic}
\bibinfo{}{Wolf, E.}
\newblock \bibinfo{}{Electromagnetic diffraction in optical systems. i. an
  integral representation of the image field}.
\newblock \emph{\bibinfo{}{Proceedings of the Royal Society of London. Series
  A. Mathematical and Physical Sciences}} \textbf{\bibinfo{}{253}},
  \bibinfo{}{349--357} (\bibinfo{}{1959}).

\bibitem{richards1959electromagnetic}
\bibinfo{}{Richards, B.} \& \bibinfo{}{Wolf, E.}
\newblock \bibinfo{}{Electromagnetic diffraction in optical systems. ii.
  structure of the image field in an aplanatic system}.
\newblock \emph{\bibinfo{}{Proceedings of the Royal Society of London. Series
  A. Mathematical and Physical Sciences}} \textbf{\bibinfo{}{253}},
  \bibinfo{}{358--379} (\bibinfo{}{1959}).

\bibitem{nasse2010realistic}
\bibinfo{}{Nasse, M.} \& \bibinfo{}{Woehl, J.}
\newblock \bibinfo{}{Realistic modeling of the illumination point spread
  function in confocal scanning optical microscopy}.
\newblock \emph{\bibinfo{}{J. Opt. Soc. Am. A}} \textbf{\bibinfo{}{27}},
  \bibinfo{}{295--302} (\bibinfo{}{2010}).
  
\bibitem{voort19893}
\bibinfo{}{Voort, H.} \& \bibinfo{}{Brakenhoff, G.}
\newblock \bibinfo{}{3-d image formation in high-aperture fluorescence confocal
  microscopy: a numerical analysis}.
\newblock \emph{\bibinfo{}{Journal of Microscopy}} \textbf{\bibinfo{}{158}},
  \bibinfo{}{43--54} (\bibinfo{}{1989}).

\bibitem{leutenegger2010analytical}
\bibinfo{}{Leutenegger, M.}, \bibinfo{}{Eggeling, C.} \& \bibinfo{}{Hell, S.}
\newblock \bibinfo{}{Analytical description of sted microscopy performance}.
\newblock \emph{\bibinfo{}{Opt. Express}} \textbf{\bibinfo{}{18}},
  \bibinfo{}{26417--26429} (\bibinfo{}{2010}).

\bibitem{chu2012single}
\bibinfo{}{Chu, K.} \& \bibinfo{}{Mertz, J.}
\newblock \bibinfo{}{Single-exposure complementary aperture phase microscopy
  with polarization encoding}.
\newblock \emph{\bibinfo{}{Optics Letters}} \textbf{\bibinfo{}{37}},
  \bibinfo{}{3798--3800} (\bibinfo{}{2012}).
  
\bibitem{hao2010effects}
\bibinfo{}{Hao, X.}, \bibinfo{}{Kuang, C.}, \bibinfo{}{Wang, T.} \&
  \bibinfo{}{Liu, X.}
\newblock \bibinfo{}{Effects of polarization on the de-excitation dark focal
  spot in sted microscopy}.
\newblock \emph{\bibinfo{}{Journal of Optics}} \textbf{\bibinfo{}{12}},
  \bibinfo{}{115707} (\bibinfo{}{2010}).
  
\bibitem{willig2007sted}
\bibinfo{}{Willig, K.}, \bibinfo{}{Harke, B.}, \bibinfo{}{Medda, R.} \&
  \bibinfo{}{Hell, S.}
\newblock \bibinfo{}{Sted microscopy with continuous wave beams}.
\newblock \emph{\bibinfo{}{Nature Methods}} \textbf{\bibinfo{}{4}},
  \bibinfo{}{915--918} (\bibinfo{}{2007}).

\bibitem{donnert2006macromolecular}
\bibinfo{}{Donnert, G.} \emph{et~al.}
\newblock \bibinfo{}{Macromolecular-scale resolution in biological fluorescence
  microscopy}.
\newblock \emph{\bibinfo{}{Proceedings of the National Academy of Sciences}}
  \textbf{\bibinfo{}{103}}, \bibinfo{}{11440--11445} (\bibinfo{}{2006}).

\bibitem{vicidomini2011sharper}
\bibinfo{}{Vicidomini, G.} \emph{et~al.}
\newblock \bibinfo{}{Sharper low-power sted nanoscopy by time gating}.
\newblock \emph{\bibinfo{}{Nature Methods}} \textbf{\bibinfo{}{8}},
  \bibinfo{}{571--573} (\bibinfo{}{2011}).

\bibitem{hell1992properties}
\bibinfo{}{Hell, S.} \& \bibinfo{}{Stelzer, E.}
\newblock \bibinfo{}{Properties of a 4pi confocal fluorescence microscope}.
\newblock \emph{\bibinfo{}{JOSA A}} \textbf{\bibinfo{}{9}},
  \bibinfo{}{2159--2166} (\bibinfo{}{1992}).

\bibitem{born1999principles}
\bibinfo{}{Born, M.}, \bibinfo{}{Wolf, E.} \& \bibinfo{}{Bhatia, A.}
\newblock \emph{\bibinfo{}{Principles of optics: electromagnetic theory of
  propagation, interference and diffraction of light}} (\bibinfo{}{Cambridge
  Univ Pr}, \bibinfo{}{1999}).

\bibitem{goodman1968fourier}
\bibinfo{}{Goodman, J.}
\newblock \bibinfo{}{Fourier optics}.
\newblock \emph{\bibinfo{}{Appendix B Sec. B}} \textbf{\bibinfo{}{3}}
  (\bibinfo{}{1968}).

\end{thebibliography}

\section*{Acknowledgments}
PX thanks Prof. Stefan W. Hell for mentoring and training on STED nanoscopy instrumentation. We thank Dr. Thomas Lawson for critical proofreading of the manuscript. This research is supported by the ``973'' Major State Basic Research Development Program of China (2011CB809101, 2010CB933901, 2011CB707502), and the National Natural Science Foundation of China (61178076).

\section*{Methods}
Home-made codes are accessible at \url{http://bme.pku.edu.cn/~xipeng/tools/STED3D.html} (replace with Nature document url once accepted). The parameters we used are listed as follows: numerical aperture of the objective lens $NA=1.4$, refractive index of the objective space $n=1.5$, the focus length of the lens $f=1.8$ mm, the excitation wavelength $\lambda_{ex}=635$ nm, the depletion wavelength $\lambda_{de}=760$ nm, and the incident excitation beam was set to be circularly polarized.

\section*{Author contribution statement}
P. X. conceived the project. H. X. accomplished the theory and program coding. H. X. and YJ. L. performed the data analysis. H. X., DY.J, and P. X. drafted the manuscript together. All authors commented and approved of the manuscript.
\section*{Competing financial interests}
The authors declare no competing financial interests.

\section*{Corresponding authors}
Correspondence to: Peng Xi.
\newpage
\section*{Figure captions}
Fig. 1  Schematic diagram of the diffraction of a high NA microscopic objective.
Fig. 2  The intensity profile of (a) excitation and (b) depletion beams on x-axis of the focal plane. Simulation intensities are plotted in red solid, polynomial approximation of Eq. (\ref{eq:excitationanddepletion}) are plotted in blue dash, and the exponential approximation is plot in green in (a).\\
Fig. 3  The relationship between the  depletion intensity and its resulted resolution. Simulation results are plotted in red solid for CW-STED in (a) and pulsed-STED (b). Then predicted resolutions of Eq. (\ref{eq:FWHM}) are plotted in blue dash in two figures, respectively.

\newpage
\center
\begin{figure}
  \centering
  \includegraphics[scale=0.5]{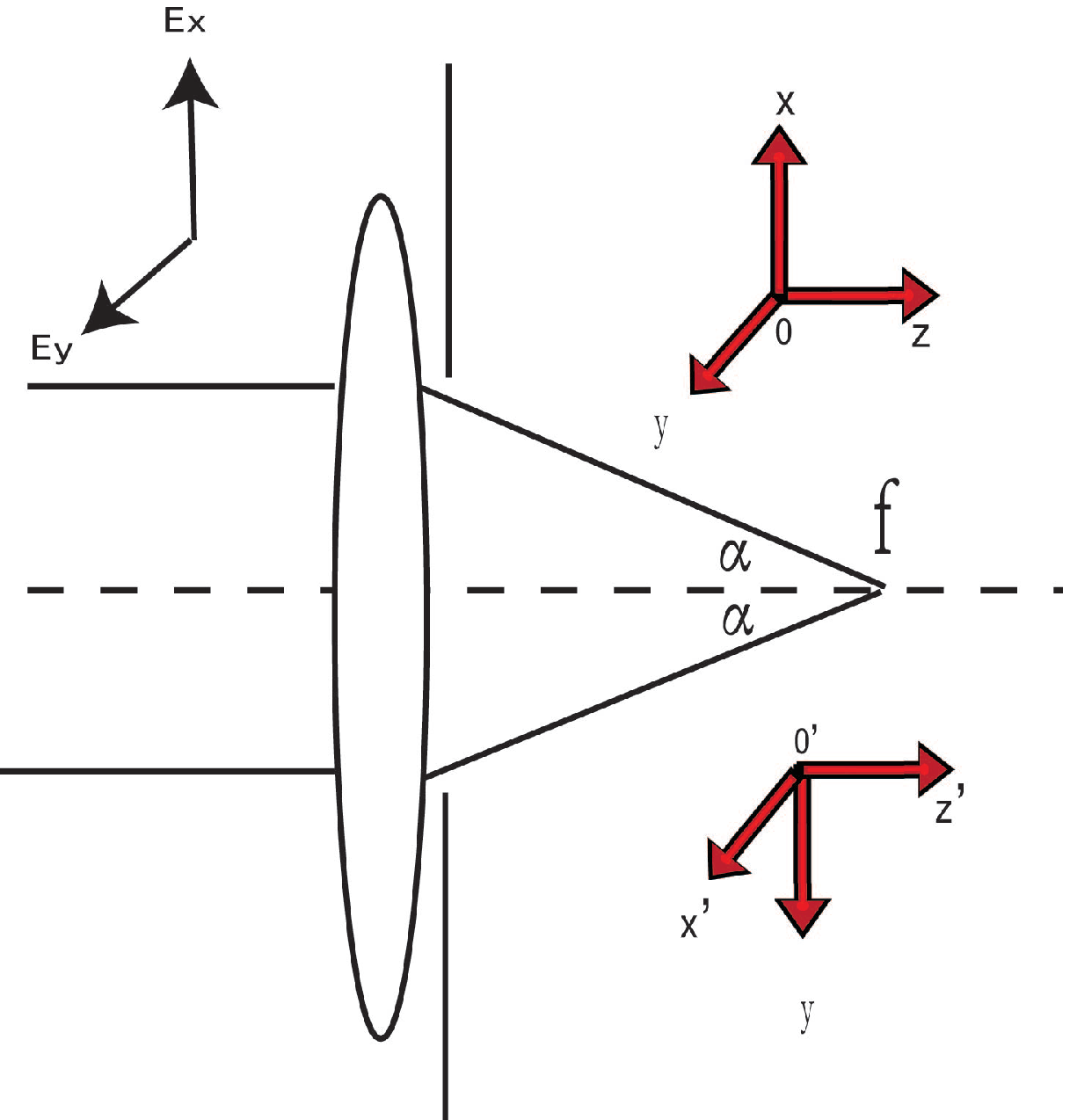}
\end{figure}
Fig. 1

\newpage

\begin{figure}
  \centering
\includegraphics[scale=1.2]{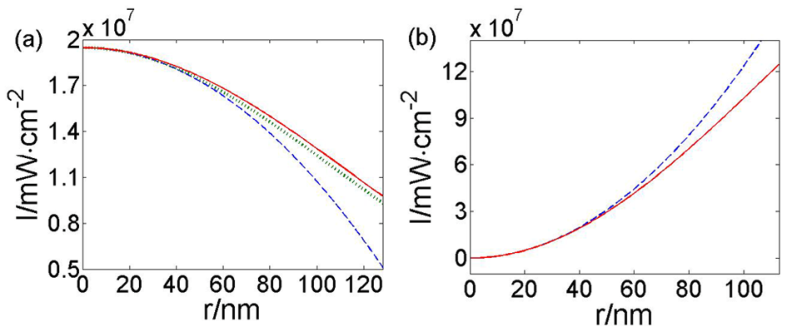}
  \label{fig:Fig4}
\end{figure}
Fig. 2

\newpage

\begin{figure}
  \centering
  \includegraphics[scale=0.25]{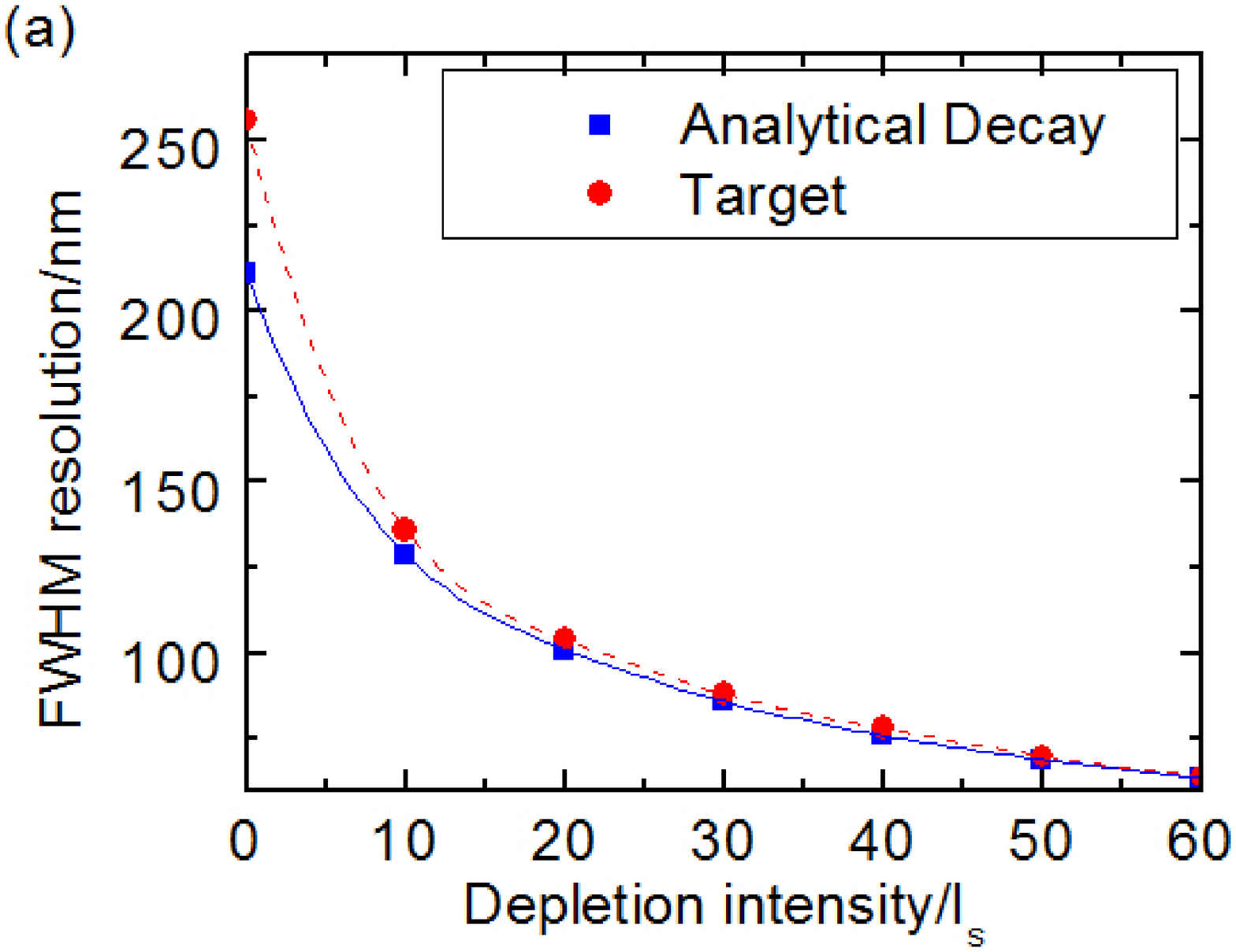}
  \includegraphics[scale=0.25]{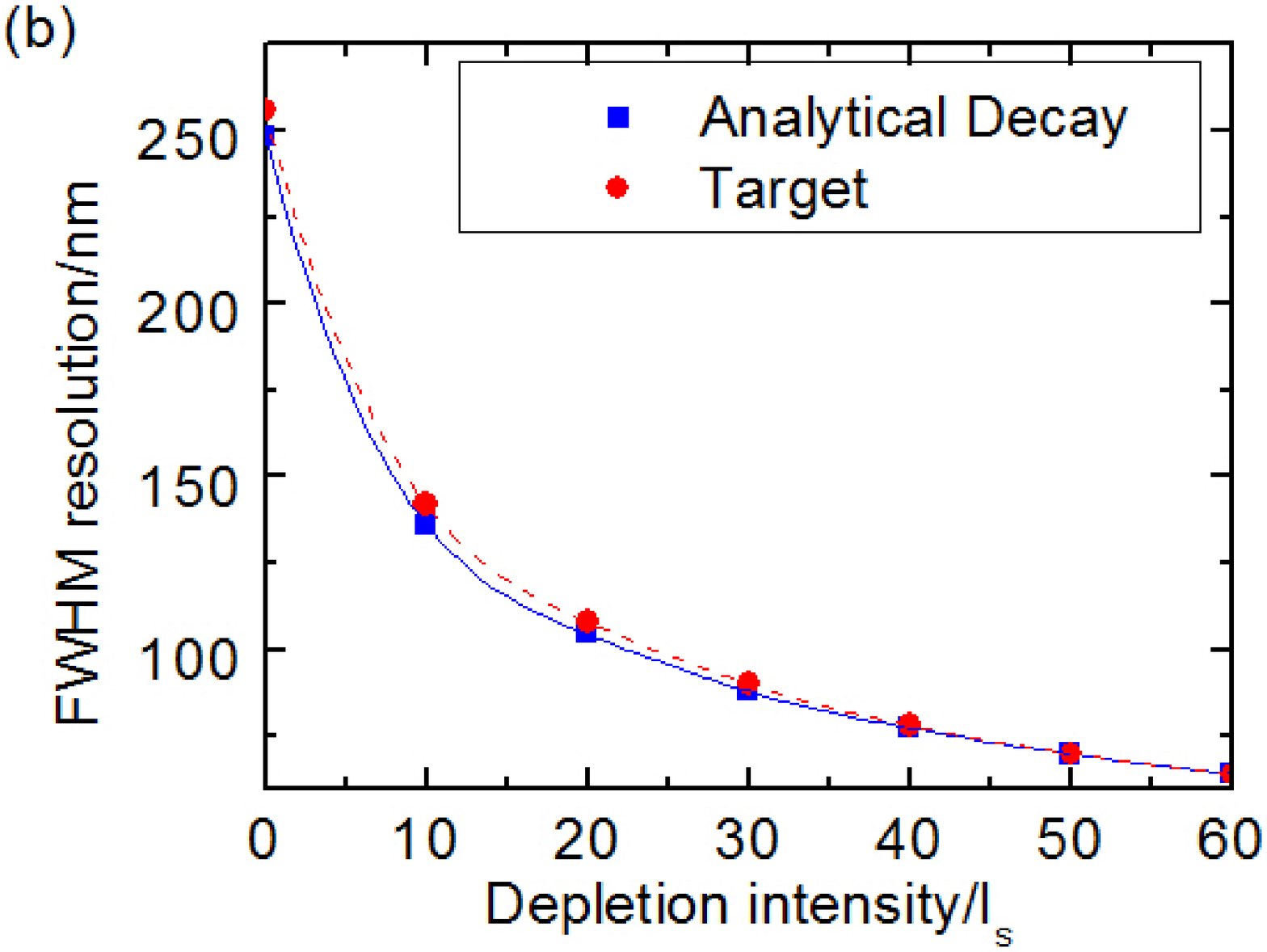}
  \label{fig:Fig5}
\end{figure}
Fig. 3

\end{document}


\title{\textbf{Analytical description of high-aperture STED resolution with 0-2$\pi$ vortex phase modulation\\
Supplementary Information}}
\date{}

\author{Hao Xie$^{1,4}$, Yujia Liu$^{1,2,3}$, Dayong Jin$^3$, and Peng Xi$^{1*}$}
\maketitle
\begin{center}
{ $^1$Department of Biomedical Engineering, College of Engineering, Peking University, Beijing 100871, China;\\
 $^2$ School of Life Sciences and Biotechnology, Shanghai Jiao Tong University, No. 800 Dongchuan Road, Shanghai 200240, China;\\
$^3$Advanced Cytometry Labs, MQphotonics Research Centre, Macquarie University, NSW 2109, Sydney, Australia;\\
$^4$Wallace H Coulter Department of Biomedical Engineering, Georgia Institute of Technology and Emory University, Atlanta, USA\\
 \url{*Email: xipeng@pku.edu.cn}
}
\end{center}

To analytically derive the field of circular polarized incident laser, it is necessary to note that in the rotated coordinations, Eq. (1) still holds with phase mask function
$e^{ i k\Phi(\theta',\phi')}=1$ for the excitation beam and $e^{ i k\Phi(\theta',\phi')}=e^{ i(\phi'+\frac{\pi}{2})}$ for the depletion beam.  It yields

\begin{equation}
\left.
\begin{aligned}
e'_{1x'}(P)&=- i A_{1y}(I_0-I_2\cos2\phi_p)\\
e'_{1y'}(P)&= i A_{1y}I_2\sin2\phi_p\\
e'_{1z'}(P)&=-2A_{1y}I_1\sin\phi_p
\end{aligned}
\right.\nonumber
\end{equation}
for excitation and
\begin{equation}
\left.
\begin{aligned}
e'_{2x'}&=A_{2y}[\mathscr{I}_1\exp( i\phi_p)+\frac{1}{2}\mathscr{I}_2\exp(- i\phi_p)-\frac{1}{2}\mathscr{I}_3\exp( i3\phi_p)]\\
e'_{2y'}&=\frac{ i A_{2y}}{2}[\mathscr{I}_2\exp(- i\phi_p)+\mathscr{I}_3\exp( i 3\phi_p)]\\
e'_{2z'}&=- A_{2y}[\mathscr{I}_4+\mathscr{I}_5\exp( i 2\phi_p)]
\end{aligned}
\right. \nonumber
\end{equation}
for depletion.

In Eq. (4) the parameters are defined as
\begin{equation}
\left.
\begin{aligned}
I_0&=a_{01}+a_{02}(k_1\rho_p)^2\\
I_1&=a_1 (k_1\rho_p)\\
I_2&=a_2 (k_1\rho_p)^2\\
\mathscr{I}_1&=b_1 (k_2\rho_p) \\
\mathscr{I}_2&=b_2 (k_2\rho_p)\\
\mathscr{I}_3&=0\\
\mathscr{I}_4&=b_{41}+b_{42} (k_2\rho_p)^2\\
\mathscr{I}_5&=b_5 (k_2\rho_p)^2
\end{aligned}
\right\}\label{eq:integration}
\end{equation}
where $a_{01}=[B(s,1,3/4)+B(s,1,5/4)]/2$,$a_{02}=[B(s,2,3/4)+B(s,2,5/4)]/8$, $a_1=B(s,2,3/4)/4$, $a_2=[B(s,2,\frac{3}{4})-B(s,2,\frac{5}{4})]/16$, $b_1=[B(s,3/2,3/4)+B(s,3/2,5/4)]/4$, $b_2=[B(s,3/2,3/4)-B(s,3/2,5/4)]/4$, $b_{41}=B(s,3/2,3/4)$, $b_{42}=B(s,5/2,3/4)/4$, $b_5=B(s,5/2,3/4)/8$.

To obtain Eq.(5) it is important to keep in mind that without phase plate $I_{dep}=D^2 a_{01}^2$. If we defined the depletion power as $D_{s}^2$ when peak fluoresce intensity into half, we also has $I_{s}=D_s^2 a_{01}^2$. So depletion later power $D^2/D_s^2$ in Eq.(4) is equivalent to peak intensity ratio $I_{dep}/I_s$ in Eq.(5). For continuous wave, $I_{ex}(\rho)\eta(I_{dep}(\rho))=I_{ex}(0)/2$, $\eta(I_{dep})=1/(1+I_{dep}/I_s)$, $I_{ex}=C^2[a_{01}^2+2(a_{01}a_{02}+a_1^2)(k_1\rho)^2]$and $I_{dep}(\rho)=D_1^2b_1^2(k_2\rho)^2$ yield
\begin{equation}
\left.
\begin{aligned}
FWHM=\frac{\lambda_{ex}}{\pi n\frac{\sqrt{-4(a_{01}a_{02}+a_1^2)}}{a_{01}}\sqrt{1+\frac{b_1^2}{2(a_{01}a_{02}+a_1^2)}
(\frac{\lambda_{ex}}{\lambda_{de}})^2\frac{I_{STED}}{I_s}}}
\end{aligned}
\right.\label{eq:FWHM1}
\end{equation}
And for continuous wave, $I_{ex}(\rho)\eta(I_{dep}(\rho))=I_{ex}(0)/2$, $\eta(I_{dep})=exp(-ln2 I_{dep}/I_s)$, $I_{ex}=C^2 a_{01}^2 exp{1+2(a_{01}a_{02}+a_1^2)/a_{01}^2}(k_1\rho)^2$ and $I_{dep}(\rho)=D_1^2b_1^2(k_2\rho)^2$ yield
\begin{equation}
\left.
\begin{aligned}
FWHM=\frac{\lambda_{ex}}{\pi n\sqrt{\frac{-2(a_{01}a_{02}+a_1^2)}{(ln 2) a_{01}^2}}
\sqrt{1+\frac{(ln 2)b_1^2}{2(a_{01}a_{02}+a_1^2)}(\frac{\lambda_{ex}}{\lambda_{de}})^2\frac{I_{STED}}{I_s}}}
\end{aligned}
\right.\label{eq:FWHM2}
\end{equation}
Then in Eq.(5) factor $\alpha$ and $\beta$ can be obtained.

\begin{figure}
  \centering
\includegraphics[width=12cm]{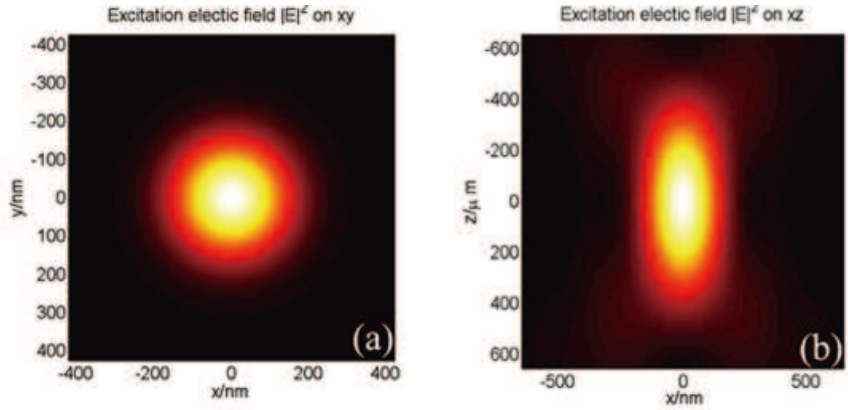}
  \caption{Excitation intensity on (a) xy plane and (b) xz plane.}\label{fig:Fig1}
\end{figure}

\begin{figure}
  \centering
  \includegraphics[width=12cm]{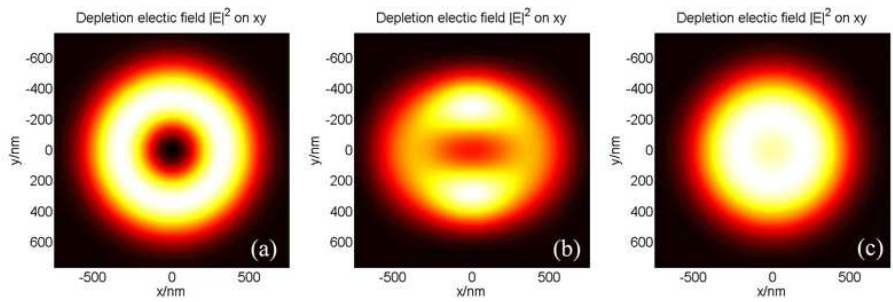}

  \caption{Intensity on the xy plane of the focus for the depletion laser with (a)left-circular polarized, (b) linear polarized, and (c) right-circular polarized laser.}\label{fig:Fig2}
\end{figure}

\begin{figure}
  \centering
  \includegraphics[width=12cm]{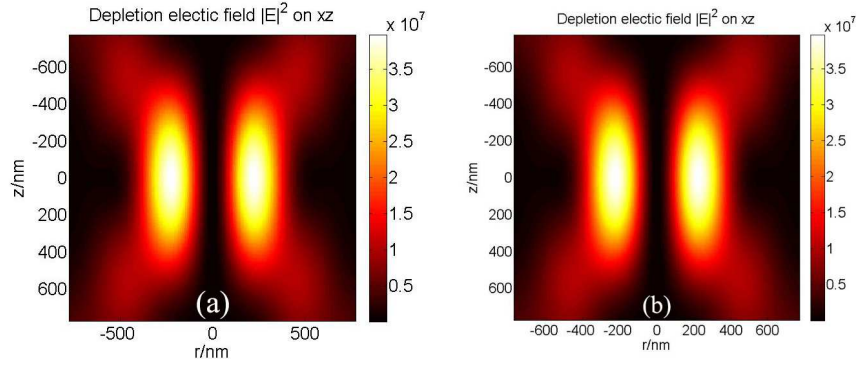}
  \caption{Intensity of the left-circular polarized depletion beam with Eq. (1) in (a) and Eq. (2) in (b)}\label{fig:Fig3}
\end{figure}

To verify our code  STED3D, first we simulated the excitation intensity on xy and xz plane in Fig. S\ref{fig:Fig1}. The FWHM is 256.8 nm and 605.7 nm respectively, in good agreement with result from PSFlab [19], thus verifying our code. Then we plotted in Fig. S\ref{fig:Fig2} the intensity on the xy plane of the focus for the depletion beam with left-circular polarized (a), linear polarized (b), and right-circular polarized laser (c). It confirms our conclusion that only the left-circular polarized depletion beam results in zero intensity in the focus. Also we plotted in Fig. S\ref{fig:Fig3} intensity of left-circular polarized depletion beam with Eq. (S2) in (b). The simulated STED distribution in both linear and circular polarization agreed well with the results of [23], and again the matching results verified our derivation.

In the discussion part we referred the saturation power $\mathscr{P}_s$ corresponds to about twice the saturation intensity $I_s$. To illustrate this, we plotted the nonlinear curves of integration depleted fluorescence efficiency and depletion intensity in Figure S4. And we also listed different factor $\xi$ with various depletion wavelengths of STED dyes in Table S1.

\begin{figure}
  \centering
  \includegraphics[width=8cm]{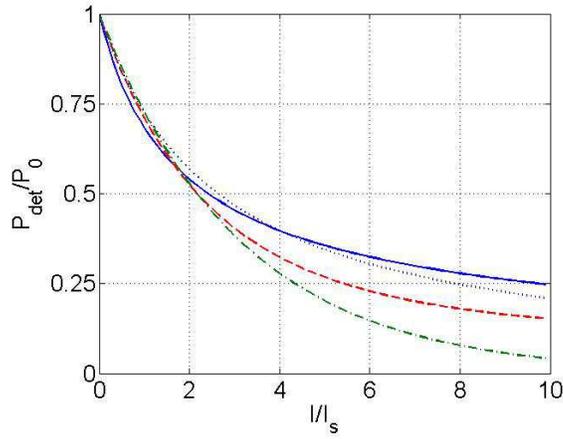}\\
    \caption{Integration depleted fluorescence efficiency and depletion intensity. Simulation results are plot in blue solid for CW-STED and red dash for pulsed- STED. Then fractal (blown dot) and exponential (green dash and dot) fits are performed for two conditions respectively.}\label{fig:Fig6}
\end{figure}
\begin{table}
\begin{tabular}{|c|c|c|c|c|}
\hline
\multirow{2}{*}{Dye}&\multicolumn{2}{|c|}{Wavelengths/nm}&Pulsed STED&CW STED\\
\cline{2-3}
&Excitation&Depletion&$\chi$&$\chi$\\
\hline
ATTO 532&470&603&0.31&0.40\\
Alexa Fluor 488, Chromeo 488&\multirow{2}{*}{488}&\multirow{2}{*}{592}&\multirow{2}{*}{0.36}&\multirow{2}{*}{0.45}\\
FITC, DY-495, Citrine&&&&\\
GFP&490&575&0.42&0.53\\
Citrine, YFP&490&600&0.35&0.44\\
Pyridine 2, RH414&554&745&0.26&0.33\\
Alexa 594, Dylight 594, ATTO 594&570&690&0.36&0.45\\
ATTO 633, ATTO 647N,&\multirow{2}{*}{635}&\multirow{2}{*}{750}&\multirow{2}{*}{0.42}&\multirow{2}{*}{0.52}\\
Abberior STAR635&&&&\\
\hline
\end{tabular}
\caption{\label{tabone}Typical values of $\beta$ for different STED systems. Dates of excitation and depletion wavelengths are from \url{http://nanobiophotonics.mpibpc.mpg.de/old/dyes/}. Numerical aperture of the lens is set to be $NA=1.4$, and $\alpha=2.15$ for pulsed STED, $\alpha=1.82$ for CW STED.}
\end{table}